\documentstyle [12pt]{article}
\begin{document}
%\draft
\newcommand{\be}{\begin{equation}}
\newcommand{\ee}{\end{equation}}
\newcommand{\bea}{\begin{eqnarray}}
\newcommand{\eea}{\end{eqnarray}}
\newcommand{\rD}{\mbox{D}}
\newcommand{\reff}{\mbox{eff}}
\newcommand{\rR}{\mbox{R}}
\newcommand{\rL}{\mbox{L}}
\newcommand{\p}{\partial}
\newcommand{\s}{\sigma}
\newcommand{\rF}{\mbox{F}}
\newcommand{\rf}{\mbox{f}}
\newcommand{\up}{\uparrow}
\newcommand{\down}{\downarrow}
\newcommand{\la}{\langle}
\newcommand{\ra}{\rangle}
\newcommand{\rd}{\mbox{d}}
\newcommand{\ri}{\mbox{i}}
\newcommand{\re}{\mbox{e}}
\newcommand{\sumnn}{\sum_{\langle jk \rangle}}
\newcommand{\rk}{\mbox{k}}

\title{Ising-model description of the SU(2)$_1$ \\
quantum critical point \\in a dimerized two-leg spin-1/2 ladder}
\smallskip
\author{Y.-J. Wang$^{a,b}$ and A. A. Nersesyan$^{a,c}$}
\maketitle
\centerline{\it $^a$The Abdus Salam International Centre for Theoretical 
Physics, } 
\centerline{\it Strada Costiera 11, 34014 Trieste, Italy}
\centerline{\it $^b$Department of Physics, Nanjing University, 210093 Nanjing, China}
\centerline{\it $^c$The Andronikashvili Institute of Physics, Tamarashvili 6, 380077 Tbilisi, 
Georgia}

\begin{abstract}
A nonperturbative analytical description of the SU(2)$_1$ quantum critical point
in an explicitly dimerized two-leg spin-1/2 Heisenberg ladder is presented.
It is shown that this criticality essentially coincides with that emerging
in a weakly dimerized spin-1 chain with a small Haldane gap.  
The approach is based on the mapping onto an SO(3)-symmetric model
of three strongly coupled quantum Ising chains. This mapping is used to
establish the correspondence between all physical fields of the spin ladder
and those characterizing the SU(2)$_1$ criticality at the infrared fixed point.
\end{abstract}
\smallskip



\bigskip

\section{Introduction}

Quantum many-particle systems in one space dimension, such as
1D interacting electrons, antiferromagnetic spin chains and ladders, 
exhibit universal properties in the low-energy limit.
The field-theoretical description of these properties 
is based on an appropriately chosen 
conformally invariant (critical) theory which is 
defined in the high-energy, ultraviolet (UV) limit and 
then deformed by
a number of perturbations consistent with the structure and symmetry
of the underlying microscopic model. While a single relevant perturbation
(with scaling dimension $d<2$) drives the model away from criticality 
towards a strong-coupling gapped phase in the infrared (IR) limit, 
the interplay between several relevant
perturbations can lead to a quantum critical point where 
conformal invariance is restored, though with a central charge $c_{\rm IR}$ 
less than that at the unstable UV  fixed point \cite{zam}.
Understanding of such phase transitions in 1D models, where powerful
nonperturbative techniques are available, is of great importance
because asymptotically exact results obtained in one space dimension can be relevant
to the issue of quantum critical points in 
higher-dimensional systems, most notably, in two-dimensional
cuprate superconductors (see e.g. \cite{vojta} and references therein).

An example of  Abelian field theory displaying
a quantum critical point was recently discussed
by Delfino and Mussardo \cite{DM}. They considered
the double-frequency sine-Gordon (DSG) model,
which is a two-dimensional Gaussian model perturbed by two relevant vertex operators
with the ratio of their scaling dimensions $d_1/d_2 = 4$, 
and showed that, upon fine tuning of its parameters, the model exhibits
an Ising criticality with central charge $c_{\rm IR}=1/2$.
In non-Abelian field theories,
the existence of quantum critical points belonging to
the universality class of SU(2)$_k$  Wess-Zumino-Novikov-Witten (WZNW) model
and
resulting from the interplay between several 
symmetry-preserving relevant perturbations
was anticipated some time ago by Affleck and Haldane \cite{aff-hald}. They argued
that a massive phase of a translationally
invariant spin-chain Hamiltonian can be driven to criticality
by an external, parity-breaking perturbation, such as an explicit dimerization.
Apparently, the explicitly dimerized even-leg spin-1/2 ladders,
which in the absence of external perturbations 
are known to possess a fully gapped excitation spectrum (see for a review 
Ref.~\cite{dagotto}), are excellent candidates for such a scenario.
A nice example of this kind has been recently given
by Martin-Delgado {\it et al} \cite{snake,snake1} 
(see also \cite{cabra})
who considered the
standard $J$-$J_{\perp}$ two-leg spin ladder $(J, J_{\perp} >0) 
$
\bea
H_{\rm stand} =  J \sum_{n}\sum_{i=1,2} %J_i (n) 
{\bf S}_{i,n} \cdot {\bf S}_{i,n+1} 
+ J_{\perp} \sum_n {\bf S}_{1,n} \cdot {\bf S}_{2,n} \;,
\label{snake}
\eea 
and modified it by making
the constituent chains dimerized with a relative phase $\pi$:
$
J \rightarrow J_{1,2} (n) = J \pm (-1)^n J'.
$
At $J = \pm J' = \frac{1}{2} J_{\perp}$ the dimerized ladder
transforms to a snake looking,
translationally invariant $S$=1/2 Heisenberg chain which is critical and 
belongs to the  SU(2)$_1$ WZNW universality class
with central charge $c = 1$ \cite{aff-hald}. Using the mapping onto a 
nonlinear sigma-model with a topological ($\theta$) term, in 
Ref.~\cite{snake} (see also \cite{totsuka1})
it was argued that, at a given $J > 0$, there should exist a critical line 
$J_{\perp} = J_{\perp} (J')$ along which the model displays the IR
properties of a single $S$=1/2 Heisenberg chain.

An analytical description of the critical points resulting from the
competition between different relevant perturbations
requires an entirely nonperturbative scheme able to correctly
identify those degrees of freedom that remain massive and the
low-energy ones that eventually become critical.
Such a scheme has been recently proposed in
Ref.~\cite{FGN-2} to tackle the Ising transition in the DSG model.
The approach was based on the mapping 
onto a 
model of two quantum Ising (QI) chains coupled by an Ashkin-Teller (AT) interaction
and also by a magnetic-field type coupling $h \s^z _1 \s^z _2$.
In the limit where the amplitude $h$ is 
considered as the largest energy scale in the problem,
the ``relative'' Ising degrees of freedom described 
by $\tau^z = \s^z _1 \s^z _2$ become locked, while the ``total'' degrees of freedom, 
described e.g. by $\s^z _1$, asymptotically decouple and can be tuned to criticality.
Universality then implies that such separation between
the fast and slow modes is valid close to the transition at arbitrary $h$.
This approach made it possible to study the Ising criticality
in a number of applications of the DSG model to physical systems \cite{FGN-2,FGN}.

In this paper, we extend this strategy to provide a consistent 
nonperturbative description of the SU(2)$_1$ criticality in the $\pi$-dimerized
two-leg spin ladder. Such a possibility naturally follows from the correspondence 
between two weakly coupled Heisenberg chains and an O(3)$\times$Z$_2$
symmetric theory of four noncritical Ising models
with order parameters $\s_i$, developed in Refs.~\cite{SNT,NT}.
The new ingredient introduced by the
explicit relative dimerization  is a perturbation
$h \prod_{i=1}^4 \s_i$ with $h \sim J'$. 
 We stress, however, that
the problem of dimerized ladders is not only different from the DSG model in that
instead of two Ising models here we have four, but most importantly because
in the adopted Ising description and all subsequent nonlocal 
(``change-of-basis'') transformations
one has to maintain the unbroken spin rotational
symmetry. For this reason, the SU(2)$_1$ quantum criticality reveals
itself in the Ising-model description in a rather nontrivial way.
The so far existing conclusion on the existence of the $c=1$ critical point 
in the $S$=1/2 staggered ladder \cite{snake} and in the related model
of the explicitly dimerized $S$=1 chain \cite{totsuka2,kn}
was reached either by mapping onto
an O(3) nonlinear sigma model which, strictly speaking, is valid for
large $S$ only, or using a perturbative (renormalization
group) approach
combined with numerical methods.
In other cases \cite{cabra}, the $c=1$ criticality 
of the dimerized ladder (\ref{snake}) was analyzed by the Abelian bosonization
in the way suitable only for the XXZ-version of the model
but not for the SU(2) symmetric case where
the explicit dimerization field is the most relevant perturbation
in the problem.
Treating our QI-chain model in the strong-coupling limit
and applying a nonlocal duality transformation \cite{KNK,KT},
below we {\em derive} the effective Hamiltonian describing
the low-energy sector of the dimerized spin ladder which 
represents a lattice spin-1/2 Heisenberg model with bond-alternating
nearest-neighbor interactions. The SU(2)$_1$ criticality is reached
when the external dimerization enforces the exchange couplings on even and odd
bonds to coincide. This mapping is then used to establish the correspondence
between the physical fields of the spin ladder and those characterizing
the SU(2)$_1$ criticality at the IR fixed point.

The paper is organized as follows. 
In Sec. II, we review the field-theoretical approach to a weakly coupled
two-leg Heisenberg ladder and show that the SU(2)$_1$ criticality
emerging upon the explicit $\pi$-phase dimerization of the constituent
chains \cite{snake} is essentially the quantum critical point appearing in
a weakly dimerized effective spin-1 chain \cite{totsuka2,kn} with a small Haldane 
gap \cite{hald-classic}. In Sec. III, we introduce an SO(3) 
symmetric quantum AT model extended to include 
an interaction $h\s_1 \s_2 \s_3$. This model, which
represents a lattice version of the field theory
discussed in Sec. II, is treated in the large-$h$ limit, and the
effective low-energy Hamiltonian displaying the
SU(2)$_1$ criticality is derived. In Sec. IV, we establish the correspondence
between the physical fields of the spin ladder and those characterizing
the SU(2)$_1$ criticality at the IR fixed point. We conclude with a
summary and discussion. The paper contains two Appendices where we provide
some details concerning the nonlocal duality transformation and the derivation
of the effective Hamiltonian in the large-$h$ limit.

\section{Spin ladder in the continuum limit}

As shown in Refs.~\cite{SNT,NT}, at $J_{\perp}\ll J$ the original lattice 
model (\ref{snake}) can be mapped onto an O(3)$\times$Z$_2$-symmetric
theory of four massive Majorana fermions, or equivalently,
four noncritical 2D Ising models.
In the continuum limit, the effective Hamiltonian density
\bea
{\cal H}_{\rm M} &=& 
-\frac{\ri v_t}{2} \left( \vec{\xi}_R \cdot \p_x  \vec{\xi}_R
- \vec{\xi}_L \cdot \p_x  \vec{\xi}_L \right) - \ri m_t \vec{\xi}_R \cdot  \vec{\xi}_L
\nonumber\\
&& -\frac{\ri v_s}{2}\left(\xi^4 _R \p_x \xi^4 _R - \xi^4 _L \p_x \xi^4 _L \right)
- \ri m_s \xi^4 _R \xi^4 _L \nonumber\\
&& +\frac{1}{2}g_1 \left( \vec{\xi}_R \cdot  \vec{\xi}_L \right)^2
+ g_2 \left( \vec{\xi}_R \cdot  \vec{\xi}_L \right) \left(\xi^4 _R \xi^4 _L  \right)
\label{ham-maj}
\eea
describes a degenerate triplet of Majorana fields, $\vec{\xi} = 
\left(\xi^1, \xi^2 , \xi^3 \right)$, and a singlet Majorana field, $\xi^4$.
The mass terms in (\ref{ham-maj}) originate from the relevant part of the interchain
interaction, $J_{\perp} {\bf n}_1\cdot{\bf n}_2 $, where
${\bf n}_{i}$ is
the staggered magnetization of the $i$-th chain:
$
(-1)^n {\bf S}_{i,n} \rightarrow {\bf n}_i (x).
$
The triplet and singlet masses are given by
$
m_t = \lambda J_{\perp}, 
$
$
m_s = -  3\lambda J_{\perp},
$
where $\lambda >0$ is a nonuniversal constant. The marginal part of the transverse 
exchange, $J_{\perp} {\bf J}_1 \cdot{\bf J}_2$, expressed in terms of the smooth
components of the local spin densities (vector ``currents'' ${\bf J}_i$), 
gives rise to the four-fermion
interaction in (\ref{ham-maj}).

While the right and left components of the vector currents
${\bf J}_i = {\bf J}_{iR} + {\bf J}_{iL}$ admit a local representation in terms
of Majorana bilinears,
\bea
{\bf I}_{\nu} &=& {\bf J}_{1\nu} + {\bf J}_{2\nu}
= - \frac{\ri}{2} \left( \vec{\xi}_{\nu} \wedge\vec{\xi}_{\nu}  \right) \;, 
\nonumber\\
{\bf K}_{\nu} &=& {\bf J}_{1\nu} - {\bf J}_{2\nu}
= \ri \vec{\xi}_{\nu}\xi^4 _{\nu} \;; ~~~~(\nu = R,L)
\label{currents}
\eea 
this is not so for the staggered fields ${\bf n}_{i}$ and $\epsilon_{i}$,
the latter being the dimerization operator defined as
$(-1)^n {\bf S}_{i,n}\cdot {\bf S}_{i,n+1}$ $\rightarrow$ $\epsilon_{i}(x)$.
However, these fields with scaling dimension 1/2
can be expressed as products of the order and disorder
parameters, $\s_a$ and $\mu_a$, of the related Ising models
(see Refs.~\cite{SNT,GNT} for more details):
\bea
&&{\bf n}^+ \sim \left(\mu_1 \s_2\s_3\mu_4,~
\s_1 \mu_2\s_3\mu_4,~  \s_1 \s_2 \mu_3 \mu_4 \right) \;,
\label{n+}\\
&&{\bf n}^- \sim \left(\s_1\mu_2\mu_3\s_4,~
\mu_1\s_2 \mu_3\s_4,~  \mu_1\mu_2\s_3\s_4\right) \;,
\label{n-}\\
&&\epsilon^+ \; \sim \mu_1\mu_2\mu_3\mu_4 \;, ~~~~
\epsilon^- \sim \s_1 \s_2\s_3\s_4 \;,
\label{dim-pm}
\eea
where ${\bf n}^{\pm} = {\bf n}_1 \pm {\bf n}_2$, $\epsilon^{\pm} = 
\epsilon_1 \pm \epsilon_2$.
Since the Majorana mass is related to the temperature of the associated Ising model
via the relation $m \sim (T-T_c)/T_c$, 
the triplet of Ising models is disordered ($m_t > 0$) while the singlet
Ising system is ordered ($m_s < 0$). As follows from Eqs.~(\ref{n+})--(\ref{dim-pm}),
this fact plays a crucial role in the behaviour of the correlation
functions. In particular, the ground state of the system is parity
symmetric ($\la \epsilon^{\pm}\ra=0$), and the
dynamical spin susceptibility $\chi''(q,\omega)$,
calculated by Fourier transforming the asymptotics of the correlator
$\la {\bf n}^{-} (x,\tau)\cdot {\bf n}^{-} (0,0)\ra$,
exhibits the existence of a coherent $S$=1 magnon peak at
$\omega^2 = (\pi - q)^2 v^2 _t + m^2 _t$.
Thus, at low energies, the standard two-leg ladder represents
a Haldane's disordered spin liquids with
${\bf n}^{-}$ playing the role of the staggered magnetization of the
effective $S$=1 chain. 
In fact, the model (\ref{ham-maj}) can be thought as a continuum low-energy 
theory of 
a spin-1 chain with a small Haldane gap, represented by a triplet of
Majorana fermions \cite{ZF,Tsv}, coupled to a noncritical Ising model:
\bea
{\cal H} &=& \frac{\pi v_t}{2} \left({\bf I}_R \cdot {\bf I}_R +
{\bf I}_L \cdot {\bf I}_L \right) + m_t {\mbox T}{\mbox r}~ \hat{\Phi}
+ g_1 {\bf I}_R \cdot {\bf I}_L
\nonumber\\
&& -\frac{\ri v_s}{2} \left( \xi^4 _R \p_x \xi^4 _R - \xi^4 _L \p_x \xi^4 _L
\right) - \ri m_s \xi^4 _R \xi^4 _L \nonumber\\
&& + g_2 {\bf K}_R \cdot {\bf K}_L
+ h \s_4 {\mbox T}{\mbox r}~ \hat{g} \;.
\label{total-ham-cont-fin}
\eea
The first term in the r.h.side of (\ref{total-ham-cont-fin}) 
is the Hamiltonian of the critical SU(2)$_2$ WZNW model
describing universal properties of the $S$=1 chain at the exactly integrable,
multicritical point \cite{TB}, ${\bf I}_{R,L}$ are
the level-2 chiral vector currents defined in (\ref{currents}),
$\hat{\Phi}$ is a 3$\times$3 matrix 
field, which is a primary field of the
SU(2)$_2$ WZNW model with scaling dimension $d_1 = 1$, 
$({\mbox T}{\mbox r}~ \hat{\Phi} = -\ri \vec{\xi} _R \cdot \vec{\xi} _L)$,
and
$\hat{g}$ is the WZNW field in the fundamental (2$\times$2) 
representation, with dimension $d_2=3/8$. The operator
\be
\epsilon_t = {\mbox T}{\mbox r}~\hat{g} = \s_1 \s_2 \s_3
\label{su2-2-dim}
\ee
is a parity-breaking, dimerization field of the triplet sector which couples to
the order parameter $\s_4$ of the singlet Ising system.
Since the latter is in the ordered phase, in the lowest order 
one can replace in Eq.~(\ref{total-ham-cont-fin}) the operators
$\s_4$  and $\xi^4 _R \xi^4 _L $ by their expectation values
and only deal with the triplet sector. Thus,
the original problem of the dimerized spin ladder reduces
to the problem of a spin-1 chain with
a small Haldane gap and a weak bond alternation:
\be
{\cal H}_{t} = \frac{\pi v_t}{2} \left({\bf I}_R \cdot {\bf I}_R +
{\bf I}_L \cdot {\bf I}_L \right)
 + \tilde{g}_1 {\bf I}_R \cdot {\bf I}_L 
+ \tilde{m}_t {\mbox T}{\mbox r}~ \hat{\Phi} + \bar{h} \epsilon_t \;,
\label{def-WZW-2}
\ee
where $\tilde{g}_1$ and $\tilde{m}_t$ are renormalized values of the 
coupling constant and triplet
mass, respectively, and
\be
\bar{h} = h \la \s_4 \ra \;. \label{new-h}
\ee
The neglected residual interaction between the singlet modes
and those low-energy triplet degrees of freedom that eventually become critical
affects the precise location of the $c=1$ critical point but
does not alter the universal properties of the transition.

\section{Generalized SO(3) quantum Ashkin-Teller \\ model and the
SU(2)$_{\bf 1}$ criticality}

Guided by the well-known correspondence between the theory of a massive Majorana
fermion and a QI model, i.e. the Ising chain in a transverse magnetic 
field \cite{Pf,LSM}, in this section we consider the following model
of three coupled QI chains
\bea
H^{(3)}_{\rm QI} &=& - \sum_{a=1,2,3} \sum_n
\left( {\cal J} \s^z _{a,n} \s^z _{a,n+1} + \Delta \s^x _{a,n}\right)\nonumber\\
&& -K \sum_{a<b} \sum_n \left( 
\s^z _{a,n} \s^z _{a,n+1} \s^z _{b,n} \s^z _{b,n+1}
+  \s^x _{a,n} \s^x _{b,n}\right)\nonumber\\
&& +\bar{h} \sum_n \s^z _{1,n} \s^z _{2,n} \s^z _{3,n} \;.
\label{3QIC-rep}
\eea
The AT interaction between the chains is  chosen to be self-dual
and parametrized by a single constant $K$.
The last term in (\ref{3QIC-rep}) couples all three chains together.

Let us first check that the lattice Hamiltonian (\ref{3QIC-rep}) is indeed 
SO(3)-symmetric
and then show that, in the continuum limit, it reduces to 
the model (\ref{def-WZW-2}). 
The hidden SO(3) symmetry of the Hamiltonian (\ref{3QIC-rep}) 
becomes manifest in the Majorana representation \cite{shankar}. 
We introduce lattice Majorana fields, $\eta_{a,n}$ and $\zeta_{a,n}$ $(a=1,2,3)$,
satisfying the anticommutation relations 
\bea
&&\{\eta_{a,n}, \eta_{b,m} \} = \{\zeta_{a,n}, \zeta_{b,m} \}
= 2 \delta_{ab}\delta_{nm} \;, \nonumber\\
&&\{\eta_{a,n}, \zeta_{b,m} \} =0 \;,
\label{anticom}
\eea
and then make use of the Jordan-Wigner 
transformations:
\be
\s^x _{a,n} = \ri \zeta _{a,n}  \eta_{a,n} \;,~~~~
\s^z _{a,n} = \ri \kappa_a \left( \prod_{m=1}^{n} 
\s^x _{a,m} \right) \zeta_{a,n} \;.
\label{c1} 
\ee
Here $\kappa_a$ are Klein factors which anticommute among themselves,
\be
\{ \kappa_a, \kappa_b\} = 2 \delta_{ab} \;,
\label{klein}
\ee
but commute with all the other operators.
Quantum disorder operators $\mu^{x,z} _{a,n+1/2}$, 
defined on the dual lattice $\{n+1/2 \}$, are given by
\be
\mu^x _{a,n+1/2} = - \ri \zeta_{a,n} \eta_{a,n+1} \;, ~~~~
\mu^z _{a,n+1/2} =   \prod_{m=1}^{n}\s^x _{a,m} \;.
 \label{c2} 
\ee
The order and disorder operators are related to each other by the duality
transformations:
\be
\s^z _{a,n} \s^z _{a,n+1} = \mu^x _{a,n+1/2} \;, ~~~~
\mu^z _{a,n-1/2}\mu^z _{a,n+1/2} =\s^x _{a,n} \;.
\label{duality-basic}  
\ee
With the definitions (\ref{c1}), (\ref{c2}),
the lattice Majorana fields 
\bea
\eta_{a,n} &=& \kappa_a \s^z _{a,n} \mu^z _{a,n-1/2}
=  \kappa_a \mu^z _{a,n-1/2}\s^z _{a,n} \;, \label{Majorana-fusion1}\\
\zeta_{a,n} &=& \ri \kappa_a \s^z _{a,n} \mu^z _{a,n+1/2}
= - \ri \kappa_a \mu^z _{a,n+1/2} \s^z _{a,n} \;, \label{Majorana-fusion2}
\eea
satisfy the required anticommutation relations (\ref{anticom}).

Using Eqs.~(\ref{c1}), we fermionize the Hamiltonian (\ref{3QIC-rep}).
The first line in (\ref{3QIC-rep}) 
immediately transforms to an O(3)-symmetric
sum of three lattice Majorana models:
$$
H_0 = \ri \sum_{n} \left({\cal J} \vec{\zeta}_{n} \cdot \vec{\eta}_{n+1} -
\Delta \vec{\zeta}_{n} \cdot \vec{\eta}_{n}  \right) \;.
$$
The AT part of $H^{(3)}_{\rm QI}$ transforms to
\be
H_{\rm AT} 
= K \sum_{a<b} \sum_n [ \left( {\zeta}_{a,n} \eta_{a,n+1}  \right) 
\left( \zeta_{b,n}  \eta_{b,n+1}  \right)\nonumber\\
+ \left( {\zeta}_{a,n} \eta_{a,n}  \right)\left( \zeta_{b,n}  \eta_{b,n}\right)] \;, 
\label{AT-via-Maj} 
\ee
and is also O(3)-symmetric because
$$
\sum_{a<b} \left(\zeta_{a,n} \eta_{a,m}\right)\left( \zeta_{b,n} \eta_{b,m}\right)
= \frac{3}{2} + \frac{1}{2}\left( \vec{\zeta}_n \cdot \vec{\eta}_m \right)^2 \;.
$$
The product of $\s^z _{1,n} \s^z _{2,n}\s^z _{3,n}$
involves products of three Majorana fields, $\eta_{1m}\eta_{2m}\eta_{3m}$
and $\zeta_{1m}\zeta_{2m}\zeta_{3m}$.
Notice that the product of the three components of a Majorana triplet,
$
\eta_1 \eta_2 \eta_3 = (1/3!) \epsilon^{abc} \eta_a \eta_b \eta_c,
$
is an  SO(3) invariant object. Therefore 
the $\bar{h}$-perturbation in (\ref{3QIC-rep}) only breaks 
the discrete subgroup Z$_2$$\subset$O(3),
thus reducing the symmetry of the model (\ref{3QIC-rep}) to
SO(3).

If the triplet of QI chains is close to the critical point and the interchain
interaction is weak,
\be
|\Delta - {\cal J}| \ll {\cal J} ~~~~{\rm and }~~~~ K,h \ll {\cal J} \;, 
\label{w-c.limit}
\ee
one can pass to the continuum limit in which the lattice operators
$\eta_{a,n}$ and $\zeta_{a,n}$ are replaced by slowly varying Majorana fields
\[
\eta_{a,n} \to \sqrt{2a_0} \eta_a(x) \;, ~~~~
\zeta_{a,n} \to \sqrt{2a_0} \zeta_a(x) \;.
\]
Here the factor $\sqrt{2}$ ensures the correct
continuum anticommutation relations,
$
\{\eta_i(x),\eta_j(y)\}= \{\zeta_i(x),\zeta_j(y) \}
= \delta_{ij}\delta(x-y). 
$
The Hamiltonian density of three decoupled QI chains then becomes
\be
{\cal H}_0 (x) \rightarrow   \left[\ri v  
\vec{\zeta} (x) \cdot \p_x \vec{\eta}(x)  - 
\ri m_t \vec{\zeta}(x) \cdot \vec{\eta} (x)\right] \label{zero}
\ee
with $v=2{\cal J} a_0$ and $m_t =2(\Delta -{\cal J})$.
A global chiral rotation of the Majorana spinors,
\be
\xi_{aR} = \frac{- \eta_a + \zeta_a}{\sqrt{2}} \;, ~~~~
\xi_{aL} = \frac{\eta_a + \zeta_a}{\sqrt{2}} \;, \label{chiral-rot1}
\ee 
transforms (\ref{zero}) to the noninteracting triplet part of the 
Hamiltonian (\ref{ham-maj}).
On the other hand,
up to irrelevant corrections, $H_{\rm AT}$ in (\ref{AT-via-Maj})
transforms  to the marginal interaction terms in 
(\ref{ham-maj}) with $g = 8K a_0$.
The correspondence between the $\bar{h}$-terms in (\ref{3QIC-rep}) and 
(\ref{def-WZW-2}) is self-evident. 

Thus, we have shown that, in the weak-coupling limit (\ref{w-c.limit}),
the 1D quantum model (\ref{3QIC-rep})
can be regarded as 
a symmetry preserving lattice counterpart
of the continuum theory (\ref{def-WZW-2}).
General universality considerations allow us to expect that,
if the field-theoretical model (\ref{def-WZW-2}) displays a certain
quantum critical behavior, this should also be a property of the
quantum lattice model (\ref{3QIC-rep}) even when its parameters are
not restricted by the condition (\ref{w-c.limit}). It is then legitimate to
consider the model (\ref{3QIC-rep}) in the strong-coupling, large-$\bar{h}$,
limit where the description of the SU(2)$_1$ criticality greatly simplifies.

Let us pass to a new set of Ising variables, 
$s^z _{1,n}$, $s^z _{2,n}$, $\tau^z _n$,
where
\be
s^z _{1,n} = \s^z _{1,n} \;, ~~~~s^z _{2,n} = \s^z _{2,n} \;, ~~~~
\tau^z _n = \s^z _{1,n} \s^z _{2,n} \s^z _{3,n} \;.
\label{new-basis}
\ee
In the original ($\s_1,\s_2, \s_3$) representation, the local (at a given
site $n$) Hilbert space of the three-chain model is spanned on the basis
vectors $|\s_1, \s_2, \s_3 \ra$ which are eigenstates of 
$\s^z _1$, $\s^z_2$ and $\s^z _3$:
$$
\s^z _a |\s_1, \s_2, \s_3 \ra = \s_a |\s_1, \s_2, \s_3 \ra \;, ~~~~
\s_a = \pm 1 \;. ~~~~~~(a=1,2,3)
$$
The new local basis $|s_1, s_2, \tau \ra$ is defined as
\bea
s^z _a |s_1, s_2, \tau \ra &=& s_a |s_1, s_2, \tau \ra \;, ~~~~(a=1,2)\nonumber\\
\tau^z |s_1, s_2, \tau \ra &=& \tau |s_1, s_2, \tau \ra \;, \nonumber
\eea
where $s_a = \s_a$, $\tau = \s_1 \s_2 \s_3$. 
Comparing matrix elements of the operators
$\s^{\alpha}_{a,n}$ $(a=1,2,3)$ in the two basises, we find the following correspondence:
\bea
&& \s^z _{1,n} = s^z _{1,n} \;, ~~~~\s^z _{2,n} = s^z _{2,n} \;, ~~~~
\s^z _{3,n} = s^z _{1,n} s^z _{2,n} \tau^z _n \;, \nonumber\\
&& 
\s^x _{1,n} = s^x _{1,n} \tau^x _{n} \;,~~~~
\s^x _{2,n} = s^x _{2,n} \tau^x _{n} \;, ~~~~
\s^x _{3,n} = \tau^x _{n} \;. \label{old-new}
\eea 
Let $\nu^{\alpha}_{1,n+1/2}$, $\nu^{\alpha}_{2,n+1/2}$, $\rho^{\alpha}_{n+1/2}$
be the disorder operators dual to $s^{\alpha}_{1,n}$, $s^{\alpha}_{2,n}$,
$\tau^{\alpha}_n$, respectively, and obeying the duality relations similar to
(\ref{duality-basic}). Under the change of basis, the original dual spins
$\mu^{\alpha}_{a,n+1/2}$ $(a=1,2,3)$ transform as follows:
\bea
&&\mu^{z}_{1,n+1/2} =\nu^{z}_{1,n+1/2} \rho^{z}_{n+1/2} \;,~~~~
\mu^{z}_{2,n+1/2} =\nu^{z}_{2,n+1/2} \rho^{z}_{n+1/2} \;,\nonumber\\  
&&\mu^{z}_{3,n+1/2} = \rho^{z}_{n+1/2} \;,\nonumber\\
&&\mu^{x}_{1,n+1/2} = \nu^{x}_{1,n+1/2} \;,~~~~
\mu^{x}_{2,n+1/2} = \nu^{x}_{2,n+1/2} \;,\nonumber\\  
&&\mu^{x}_{3,n+1/2} = \nu^{x}_{1,n+1/2}\nu^{x}_{2,n+1/2}\rho^{x}_{n+1/2} \;.
\label{old-new-dual} 
\eea
It is easy to check that the new pairs of mutually dual operators,
$(s_1, \nu_1)$, $(s_2, \nu_2)$ and $(\tau, \rho)$, satisfy 
the same algebra, Eqs.~(\ref{Majorana-fusion1}), (\ref{Majorana-fusion2}),
as the original operators $(\s_a, \mu_a)$ $(a=1,2,3)$.

In terms of the new variables, the Hamiltonian (\ref{3QIC-rep}) reads:
\bea
H^{(3)}_{\rm QI} &=& 
- ({\cal J}+K) \sum_n \left(s^z _{1,n} s^z _{1,n+1} +  s^z _{2,n} s^z _{2,n+1} 
+ s^z _{1,n} s^z _{1,n+1}s^z _{2,n} s^z _{2,n+1} \right) \nonumber\\
&& - K\sum_n \left( s^x _{1,n} + s^x _{2,n} +
s^x _{1,n} s^x _{2,n} \right)\nonumber\\
&& + \bar{h} \sum_n\tau^z _n 
- \Delta \sum_n 
\left(s^x _{1,n} + s^x _{2,n}+1 \right)\tau^x _n 
\nonumber\\ 
&& +\frac{{\cal J}}{2} \sum_n s^z _{1,n} s^z _{1,n+1}s^z _{2,n} s^z _{2,n+1}
\left( \tau^z _n - \tau^z _{n+1} \right)^2 \nonumber\\ 
&& +\frac{K}{2} \sum_n \left( s^z _{1,n} s^z _{1,n+1} +  s^z _{2,n} s^z _{2,n+1}
 \right)\left( \tau^z _n - \tau^z _{n+1} \right)^2 \;.
\label{ham-new-rep}
\eea
The advantage of this representation of $H^{(3)}_{\rm QI}$ becomes transparent
in the strong-coupling
limit: 
\be
\bar{h} \gg {\cal J},\Delta, K \;.\label{s-c.limit}
\ee
In the zeroth-order approximation, the Hamiltonian (\ref{ham-new-rep})
describes a collection of noninteracting $\tau$-spins in a strong magnetic
field, 
with a large energy gap $2\bar{h}$ separating 
the fully polarized ground state from the excited states with one spin flip.
Thus the $\tau$ degrees of freedom are ``fast'' and can therefore be integrated
out to produce an effective Hamiltonian for the low-energy, ($s_1,s_2$) part
of the spectrum.

Using a unitary transformation in the form of a $1/\bar{h}$ expansion, 
we project $H^{(3)}_{\rm QI}$ onto the lowest-energy
state ($\tau^n = -1,~\forall n$) 
of the zeroth-order Hamiltonian, $H_0 = \bar{h}\sum_n\tau^z _n$, and thus obtain
the effective model of two coupled QI chains,
$H_{\rm eff}[s_1,s_2]$.
Apparently, the symmetry of $H_{\rm eff}[s_1,s_2]$
might appear only as Z$_2$$\times$Z$_2$, for $H^{(3)}_{\rm QI}$ in (\ref{ham-new-rep})
contains only $s^x _{1,2}$ and $s^z _{1,2}$. The hidden SU(2) symmetry of the
effective model is revealed by the nonlocal duality transformation first introduced
by Kohmoto, den Nijs and Kadanoff (KNK) in their study of
the 2D AT model \cite{KNK}, and later on employed by Kohmoto and Tasaki
in their analysis of the bond-alternating $S$=1/2 chain \cite{KT}.
This transformation, which is briefly reviewed in Appendix A,
establishes an important correspondence between two combinations
of the variables $s^{\alpha}_1$, $s^{\alpha}_2$ $(\alpha=z,x)$  
and local SU(2) invariants written in terms of new spin-1/2 operators
${\bf T}_{n}$:
\bea
s^z _{1,n} s^z _{1,n+1} +  s^z _{2,n} s^z _{2,n+1} 
+ s^z _{1,n} s^z _{1,n+1}s^z _{2,n} s^z _{2,n+1}
&\rightarrow&  - 4 {\bf T}_{2,n}\cdot {\bf T}_{2n+1}\label{su2-inv1} \;, \\
s^x _{1,n} + s^x _{2,n} + s^x _{1,n} s^x _{2,n}
&\rightarrow& - 4 {\bf T}_{2n-1}\cdot {\bf T}_{2,n} \;, \label{su2-inv2}
\eea
where $\{2n\}$ and $\{2n+1\}$ denote even and odd sublattices 
of a new lattice with $2N$ sites. Notice that the duality transformation
that maps the l.h.sides of Eqs.~(\ref{su2-inv1}) and (\ref{su2-inv2})
onto each other is equivalent to a translation by one lattice spacing
on the lattice where the spin-1/2 operators ${\bf T}_{n}$ are defined.

To the accuracy $O(1/\bar{h}^2)$, the low-energy effective Hamiltonian
is found to be (see Appendix B)
\bea
H_{\rm eff}[s_1,s_2]&=& \sum_{n=1}^{N} \left( J_1 {\bf T}_{2,n}\cdot {\bf T}_{2n+1}
+ J_2 {\bf T}_{2n-1}\cdot {\bf T}_{2,n} \right)\nonumber\\
&& +J_3 \sum_{n=1}^{N} \left( {\bf T}_{2n-1} \cdot {\bf T}_{2n+1}
+ {\bf T}_{2,n}\cdot {\bf T}_{2n+2} \right) \;,
\label{eff-final}
\eea
where
\bea
J_1 = 4 \left[{\cal J} + K - 
\left(\frac{\Delta}{\bar{h}}\right)^2 \left({\cal J} + 2K \right)  \right] \;,
\nonumber\\
J_2 = 4 \left(K + \frac{\Delta^2}{\bar{h}}\right) \;, ~~~~
J_3 = 4 \left(\frac{\Delta}{\bar{h}}\right)^2 K \;.
\label{J1-J2-J3}
\eea
$H_{\rm eff}[s_1,s_2]$ is manifestly SU(2) invariant 
and describes a spin-1/2
chain with bond-alternating nearest-neighbor interactions $J_1$ and $J_2$
and the next-nearest-neighbor interaction $J_3$. 
The SU(2)$_1$ critical regime
occurs when the bond alternation vanishes:
$J_1 = J_2$. Notice that the condition
$\Delta/\bar{h} \ll 1$ ensures that the frustrating interaction $J_3$ remains
irrelevant at the transition. Eqs.~(\ref{eff-final}), 
(\ref{J1-J2-J3}) constitute the central result of our work.

It is worth stressing here that 
the equivalence between two coupled critical QI chains and a bond-alternating
SU(2)-symmetric $S$=1/2 Heisenberg model necessarily implies that the AT
interchain interaction is of the order of the cutoff and
fine tuned in such a way that all three terms
in the l.h.sides of Eqs.~(\ref{su2-inv1}), (\ref{su2-inv2})
have the same amplitude. This
is a manifestation of the well-known fact, implicitly present in earlier studies
of the 2D AT model \cite{KNK}, that there exists no free-fermion
Majorana representation of the critical SU(2)$_1$ WZNW model. At the point
where the effective $N$=2 AT model $H_{\rm eff}[s_1,s_2]$ is self-dual,
the $S$=1/2 Heisenberg chain becomes translationally invariant and critical.
In the 2D $N$=2 AT model, this is the point where the $c=1$ line of Gaussian critical
points with continuously varying exponents bifurcates
into two Ising critical lines \cite{KNK,ditz}.

In the lowest order, the critical value of $\bar{h}$
is given by $\bar{h}_c = \Delta^2/{\cal J}$. With $\Delta/\bar{h}$ being the small
parameter in the expansions (\ref{J1-J2-J3}), the ratio 
$\Delta/{\cal J} \sim \bar{h}_c /\Delta \gg 1$, implying that the 
$c=1$ criticality is reached if the three QI chains in (\ref{3QIC-rep})
are strongly disordered. This is a direct consequence of 
the imposed condition (\ref{s-c.limit}) which prevents us to 
establish a quantitative correspondence 
between the parameters of
the field-theoretical models
(\ref{total-ham-cont-fin}), (\ref{def-WZW-2}) and the 
lattice model (\ref{3QIC-rep}), and, in particular, determine
the critical line. However,
universality arguments allow us to expect that the asymptotic
decoupling between the $\tau$-degrees of freedom, which become frozen
at the SU(2)$_1$ transition, and the $(s_1,s_2)$ degrees of freedom that
describe an effective critical $S$=1/2 Heisenberg chain is a general
property of the quantim AT model (\ref{3QIC-rep}) with an arbitrary
value of $\bar{h}$. This means that, even when the condition
(\ref{s-c.limit}) is released, the low-energy sector of the model
(\ref{3QIC-rep}) will still be described by an effective Hamiltonian
(\ref{eff-final}), although the determination of its parameters
will remain as a complicated, yet unresolved, part of the problem.

Despite this shortcoming, it is possible to extract the scaling of the
critical line in the weak-coupling limit from general considerations 
and compare the results with the already existing ones. Let us go back
to the model (\ref{def-WZW-2}). At small $\bar{h}$ and $\tilde{m}_t$,
the scaling law for the critical line $\bar{h}_c = \bar{h}_c (\tilde{m}_t)$ 
simply follows from the comparison of
the mass gaps, $\sim \bar{h}^{8/13}$ and $\sim \tilde{m}_t$,
generated by the $\bar{h}$- and $\tilde{m}_t$-perturbations 
independently \cite{kn}:
\be
\bar{h}_c \sim [\tilde{m}_t]^{13/8} \sim J^{13/8}_{\perp} \;. \label{c=1.crit.line}
\ee
For the standard ladder where $|m_s| \simeq  3 |m_t| \sim J_{\perp}$,
one has $\la \s^z _4 \ra \sim J^{1/8}_{\perp}$, and the relation
(\ref{c=1.crit.line}) translates to 
\be
h_c \sim \la \s^z _4 \ra^{-1} J_{\perp}^{13/8}
\sim J_{\perp}^{3/2} \;, \label{c=1.critline-stanladd}
\ee
in agreement with the result of Ref.~\cite{snake1}.

Returning to the adopted mean-field treatment of the singlet Ising component,
it is worth noticing here that the description of the lowest-energy part
of the spectrum in terms of the effective spin-1/2 Hamiltonian (\ref{eff-final})
is valid not only at the SU(2)$_1$ transition but also at small deviations
from criticality, provided that  the associated (dimerization) mass gap
$m_{\rm dimer} \sim |\bar{h} - \bar{h}_c |^{2/3}$ is much smaller than
the singlet mass $|m_s|$.

\section{UV-IR transmutation of physical fields}

Let us now find out how the physical fields,
characterizing the spin ladder in the UV limit, are transformed
at the IR fixed point where the system becomes SU(2)$_1$ critical.

We start with the current operators ${\bf J}_{1,2}$ which, according to
Eqs.~(\ref{currents}), are expressed in terms of the Majorana fields
$\vec{\xi}$ and $\xi^4$. Let us first consider the total current,
$$
{\bf I} = {\bf J}_{1} + {\bf J}_{2} =
- \frac{\ri}{2}\left(\vec{\xi}_R \wedge \vec{\xi}_R
+ \vec{\xi}_L \wedge \vec{\xi}_L  \right),
$$
which is fully determined in the 
triplet sector of the model and, in the UV limit, actually represents the level-2 
vector current of the SU(2)$_2$ WZNW model given by the first term
in Eq. (\ref{def-WZW-2}). This current is nothing but
the smooth part of the magnetization
of the related critical S=1 Heisenberg chain.
The $x$-component of this current is
\be
I^x (x) = - \ri \left[ \xi^2 _R (x) \xi^3 _R (x)
+ \xi^2 _L (x) \xi^3 _L (x)\right] \;.
\label{Ix-cont}
\ee
Making the chiral rotation (\ref{chiral-rot1}), we can define a local lattice
operator
\be
I^x _n = - \frac{\ri}{2} \left(\eta_{2,n} \eta_{3,n} + \zeta_{2,n} \zeta_{3,n}  \right)
\rightarrow a_0 I^x (x) \;,
\label{Ix-lattice}
\ee
which transforms back to (\ref{Ix-cont}) in the continuum limit.
Using definitions (\ref{Majorana-fusion1}), (\ref{Majorana-fusion2}),
together with the ``change-of-basis'' transformations (\ref{old-new}),
(\ref{old-new-dual}), we obtain:
\bea
I^x _n &=& \frac{\ri \kappa_2 \kappa_3}{2}
\s^z _{2,n} \s^z _{3,n} \left(\mu^z _{2,n+1/2} \mu^z _{3,n+1/2} 
- \mu^z _{2,n-1/2} \mu^z _{3,n-1/2} \right)\nonumber\\
&=& \frac{\ri \kappa_2 \kappa_3}{2} \tau^z _n s^z _{1,n}
\left( \nu^z _{2,n+1/2} - \nu^z _{2,n-1/2} \right) \;.
\label{Ix-intermed}
\eea
Since $\tau^z _n$ is a noncritical field, it can be replaced by its nonzero expectation
value. Using formulas (\ref{spins-KT}) of 
Appendix A, we express $I^x _n$ in terms
of the spin operators of the effective critical $S$=1/2 Heisenberg chain
(\ref{eff-final}):
\be
I^x _n = \ri \kappa_2 \kappa_3 \kappa_1 \la \tau^z \ra
\left( T^x _{2n} + T^x _{2n +1} \right) \;.
\ee

In the continuum limit
\bea
{\bf T}_m &\rightarrow& (a_0 / 2) {\bf T}(x) \;, \nonumber\\
{\bf T}(x) &=& {\bf J} (x) + (-1)^m {\bf n} (x) \;,
\label{T-cont.lim}
\eea
where ${\bf J} = {\bf J}_R + {\bf J}_L$ and ${\bf n}$ are the 
smooth and staggered parts of the local magnetization ${\bf T}(x)$. In (\ref{T-cont.lim})
we took into account the fact that, by the KNK construction
(see Appendix A), the lattice spacing of the effective $S$=1/2 Heisenberg chain
is $a_0/2$. Keeping only relevant operators, we find that 
$I^x (x)$, being the level-2 current in the UV limit,
transforms to 
%the level-1 current at the critical point according to the following
%correspondence:
\be
I^x (x) \rightarrow \ri \kappa_2 \kappa_3 \kappa_1 \la \tau^z \ra
\left[ J^x (x) - (a_0/2) \p_x n^x (x) + \cdots\right], \;.
\label{Ix-Jx}
\ee
where $\cdots$ stand for irrelevant corrections.
Quite similarly we find that
\bea
I^y (x) &\rightarrow&  \ri \kappa_3 \kappa_1 \kappa_2 
\la \tau^z \ra  
\left[ J^y (x) - (a_0/2) \p_x n^y (x) + \cdots\right] \;, \label{Iy-Jy}\\
I^z (x) &\rightarrow& %\kappa_1 \kappa_2 \kappa_2 \kappa_1 
\left[ J^z (x) - (a_0/2) \p_x n^z (x) + \cdots\right] \;. \label{Iz-Jz}
\eea
Since $\left(\kappa_1 \kappa_2 \kappa_3 \right)^2 = -1$, we can replace
$\kappa_1 \kappa_2 \kappa_3$ by $\ri$.
In the leading order in $1/\bar{h}$, $\la \tau^z \ra = - 1$, and we arrive at
the following correspondence
\be
{\bf I} (x) \rightarrow {\bf J}(x) - C a_0 \p_x {\bf n} (x)  \;,
\label{I-J.final}
\ee
($C$ being a nonuniversal constant),
which shows that the level-2 vector current ${\bf I}$, originally defined
at the UV fixed point, transforms in the IR limit not only to the level-1 vector current
${\bf J}$ but also acquires a nonholomorphic piece  represented
by the operator $\p_x {\bf n}$ with scaling dimension 3/2 and conformal spin 1.

One can similarly show that the level-2 axial vector current 
(i.e. the spin current of the critical S=1 chain), 
defined at the UV fixed point as
$$
{\bf I}_5 = - \frac{\ri}{2}\left(\vec{\xi}_R \wedge \vec{\xi}_R
- \vec{\xi}_L \wedge \vec{\xi}_L  \right),
$$
transforms to
\be
{\bf I}_5 (x) \rightarrow {\bf J}_5(x) + C a_0 \p_t {\bf n} (x)
\label{I-5}
\ee
where ${\bf J}_5 = {\bf J}_R - {\bf J}_L$ is the axial vector current
of the SU(2)$_1$ WZNW model (i.e. the spin current of the S=1/2 Heisenberg chains).
Notice that the constant $C$ in (\ref{I-5}) is the same as in (\ref{I-J.final}).
As a result, both at the UV and IR fixed points, the currents ${\bf I}$ and ${\bf I}_5$
satisfy the continuity equation: 
$$
\p_t {\bf I} + \p_x {\bf I}_5 = 0.
$$
Since the holomorphic property of both level-2 currents is lost in the IR limit,
the second equation $\p_x {\bf I} + \p_t {\bf I}_5 = 0$ is no longer valid.

According to Eq.~(\ref{currents}), the relative current
${\bf K} = {\bf J}_1 - {\bf J}_2$ involves the singlet Majorana field
$\xi^4$ which remains massive across the transition. Therefore, the current
${\bf K}$ represents a short-ranged field whose correlations decay exponentionally
over the length scale $\xi_s \sim v_s /|m_s|$.

Let us now turn to the staggered fields ${\bf n}^{\pm}$ and $\epsilon^{\pm}$
which are defined at the UV fixed point by Eqs.~(\ref{n+})--(\ref{dim-pm}).
Since ${\bf n}^+$ and $\epsilon^+$ are proportional to $\mu_4$ and since the forth
Ising model is ordered $(m_s < 0,$$~\la \mu_4 \ra = 0)$, these two fields
remain short-ranged at the transition. 
On the other hand, with $\s_4$ replaced by its nonzero expectation
value, ${\bf n}^-$ and $\epsilon^-$ transform to the staggered magnetization
and dimerization field of the effective $S$=1 chain 
\bea
{\bf n}^-  &\sim&
\left(\s_1 \mu_2 \mu_3, ~\mu_1 \s_2 \mu_3,~ \mu_1 \mu_2 \s_3  \right) \;,
\label{eff-N}\\
\epsilon^-  &\sim& \epsilon_t = \s_1 \s_2 \s_3 \;.
\label{eff-dim}
\eea

Adopting a symmetric lattice definition, for $(n^-)^x$ we obtain:
\be
(n^-)^x _n = \frac{1}{2} \s_{1,n} 
\left( \mu^z _{2,n+1/2} \mu^z _{3,n+1/2}
+ \mu^z _{2,n-1/2} \mu^z _{3,n-1/2} \right) \;. \label{Nx-latt}
\ee
Under transformations (\ref{old-new}), (\ref{old-new-dual}) and 
formulas (\ref{spins-KT}), we find that
$(n^-)^x _n$ becomes
\bea
(n^-)^x _n &=& \frac{1}{2} s^z _{1,n}\left(\nu^z _{2,n+1/2} - \nu^z _{2,n+1/2}\right)
\nonumber \\ 
&=& \frac{1}{2} \kappa_1 \left( T^x _{2,n} - T^x _{2n-1}\right) \;.
\nonumber
\eea
Passing then to the continuum limit, with (\ref{T-cont.lim}) taken into account,
we obtain:
\be
(n^-)^x (x) \sim \kappa_1 n_x \;, \label{Nx-nx}
\ee
where ${\bf n}$ is the staggered magnetization of the critical $S$=1/2 chain.
Similarly
\be
(n^-)^y (x) \sim \kappa_2 n_y \;, 
~~~~(n^-)^z (x) \sim \ri \kappa_1 \kappa_2 \la \tau^z \ra n_x \;.
\label{n-yz}
\ee
The Klein factors can be identified with Pauli matrices:
$$
\kappa_1 = \tilde{\tau}^x \;, ~~~~\kappa_2 = \tilde{\tau}^y \;, ~~~~
\ri \kappa_1 \kappa_2 = - \tilde{\tau}^z \;.
$$
Thus, in the leading order in $1/\bar{h}$ $~\left(\la \tau^z \ra = -1
\right)$, we arrive at the following correspondence:
\be
(n^-)^{\alpha} (x) \sim \tilde{\tau}^{\alpha} n^{\alpha} \;.
~~~~(\alpha=x,y,z)
\label{Nn-final} 
\ee
The presence of the Klein factors in (\ref{Nn-final}) should not
be confusing. These factors drop out of the Hamiltonians (\ref{3QIC-rep})
and (\ref{eff-final}) and have no dynamics.
Moreover, due to the unbroken SU(2) symmetry,
$\la n^{\alpha} (x,\tau) n^{\beta} (0,0) \ra = \frac{1}{3} 
\delta^{\alpha\beta} \la {\bf n} (x,\tau) \cdot {\bf n} (0,0)\ra$,
the Klein factors drop out of the correlation functions as well.
Thus, at the SU(2)$_1$ critical point, the relative staggered
magnetization of the spin ladder transforms to the staggered magnetization
of the effective $S$=1/2 chain:
\be
{\bf n}^- \rightarrow {\bf n} \;. \label{Nn-final1} 
\ee

It may seem from formulas (\ref{Ix-Jx}), (\ref{Iy-Jy}) and (\ref{n-yz})
that taking into account $1/\bar{h}$ corrections to $\la \tau^z\ra = -1$
would break SU(2) invariance 
of the UV-IR transmutation rules (\ref{I-J.final}), (\ref{Nn-final1}).
The point is that the adopted ``minimal''
choice of local lattice operators (see Eqs.~(\ref{Ix-lattice}), (\ref{Nx-latt})),
although being consistent with the continuum representation of the corresponding
fields, is not unique. Other lattice regularizations of the field operators
may be equally good in this respect
but differ in the $\bar{h}$-dependence of the prefactors and
subleading corrections at the critical point. 
With the manifestly SU(2) invariant correspondence (\ref{I-J.final}), (\ref{Nn-final1})
firmly established in the leading order, we can only state here that
these corrections are not universal and, as already
explained in the preceeding section, cannot be addressed within this approach.

Finally, we consider transmutation of the dimerization field $\epsilon^-$.
From (\ref{eff-dim}) and (\ref{old-new}) it follows that
\be
\epsilon^-  \sim \tau^z \sim - I \;.
\label{rel-dimer-I}
\ee
However, $\epsilon^-$ couples
to the amplitude $\bar{h}$ of the external dimerization whose variation
leads to the deviation from criticality. The leading 
SU(2)-symmetric, parity-breaking, relevant
perturbation
at the SU(2)$_1$ criticality is the dimerization field 
of the effective $S$=1/2 Heisenberg chain:
$(-1)^n {\bf T}_n \cdot {\bf T}_{n+1} \rightarrow \epsilon(x)$. 
Therefore, 
the improved version of the
formula (\ref{rel-dimer-I}), which includes a strongly fluctuating
correction to the identity operator $I$, should read
\be
\epsilon^- (x) \sim - I + \epsilon (x) \;.
\label{rel-dimer.correct}
\ee
Since the scaling dimension of $\epsilon$ is 1/2, the behavior of
$\la \epsilon \ra$ close to the critical point is given by
\be
|\la \epsilon^- \ra|_{\bar{h}} =
|\la \epsilon^- \ra|_{\bar{h}_c} + {\rm const.}\, |\bar{h} - \bar{h}_c|^{1/3} \, 
{\rm sign} (\bar{h} - \bar{h}_c) \;.
\label{epsilon-close-to-crit}
\ee
Thus, the average dimerization remains finite at the transition
but has an infinite slope.

Let us now draw a physical picture emerging from the obtained results.
In the limit of a translationally
invariant ladder, $J'/J_{\perp} \rightarrow 0$, the lowest-energy part of
the spin fluctuation spectrum displays a single coherent $S$=1 massive magnon formed
due to confinement of the originally massless $S$=1/2 spinons of
individual chains. In the opposite limit of two decoupled bond-alternating 
spin-1/2 chains,
$J_{\perp}/J' \rightarrow 0$, 
coherent massive magnons are still present in the spectrum but are of a different
nature: these represent two independent sets of triplet states formed
by a soliton, antisoliton and the first breather of the two
corresponding $\beta^2 = 2\pi$
sine-Gordon models (see e.g. Ref.~\cite{GNT}, chapter 22).
The crossover between
these two extreme cases involves a critical point where the spectrum
is entirely incoherent and consists of pairs of $S$=1/2 spinons
of the effective $S$=1/2 Heisenberg chain. (The behavior of the current-current
correlation function $\la {\bf I}(x,\tau) {\bf I}(0,0) \ra$ at the transition
will be modified due to the admixture of the nonholomorphic operator in
Eq. (\ref{I-J.final})). As follows from (\ref{eff-final}), (\ref{J1-J2-J3}),
close to the transition, the low-energy properties of the system
are those of a {\em single}, weakly dimerized spin-1/2 Heisenberg chain.
(The fact that the average relative dimerization of the original
ladder, $\la  \epsilon^- \ra$, stays finite across the transition implies
that higher-energy degrees of freedom which remain massive at the critical
point are also dimerized.)

The two massive phases separated by the $c=1$ critical point
differ in the sign of the explicit dimerization of the effective
$S$=1/2 chain (\ref{eff-final}) and
are characterized by two different string order 
parameters \cite{den}, each of them
being nonzero only in one phase and vanishing in the other. For the spin-1/2
alternating Heisenberg chain, the nonlocal string order parameter
associated with the breakdown of a hidden Z$_2$$\times$Z$_2$ symmetry
was first considered by Kohmoto and Tasaki \cite{KT} and Hida \cite{hida}.
Its representation in terms of the Ising order and disorder parameters
for two-leg spin-1/2 ladders was introduced in Refs.~\cite{SNT,NT}.
The two different string order parameters of the alternating chain
(\ref{eff-final}) are defined as follows (due to the unbroken SU(2)
symmetry, it is sufficient to consider only the $x$-components of the 
correspoding operators):
\bea
O^x _{2k, 2n-1} &=& \exp \left( \ri \pi \sum_{j=2k}^{2n-1} T^{x}_j \right) \;,
\label{string-1}
\nonumber\\
O^x _{2k+1, 2n} &=& \exp \left( \ri \pi \sum_{j=2k+1}^{2n} T^{x}_j \right) \;.
\label{string-2}
\eea
Using relations (\ref{bilin-KT1}) and (\ref{bilin-KT2}) given in Appendix
A, we find that
\be
O^x _{2k, 2n-1} = s^z _{1,k} s^z _{1,n} \;, ~~~~
O^x _{2k+1, 2n} = \mu^z _{2,k+1/2} \mu^z _{2,n+1/2} \;, \label{string-ising}
\ee
where $(s^z _{1,k}, \mu^z _{1,k+1/2})$ and $(s^z _{2,k}, \mu^z _{2,k+1/2})$
are pairs of the order and disorder operators of two QI chains representing
the effective spin-1/2 Heisenberg Hamiltonian (\ref{eff-final})
(see Sec. III). As follows from (\ref{J1-J2-J3}), close to the
transition, $J_1 - J_2 \sim \bar{h} - \bar{h}_c$. Since
at $\bar{h} < \bar{h}_c$ the two QI chains are disordered,
one finds that
\be
\la O^x _{2k+1, 2n} \ra \neq 0 \;, ~~~~\la O^x _{2k, 2n-1} \ra = 0 \;.
\ee
At $\bar{h} > \bar{h}_c$ these chains are ordered implying that
in this case
\be
\la O^x _{2k+1, 2n} \ra = 0 \;, ~~~~\la O^x _{2k, 2n-1} \ra \neq  0 \;.
\ee
As shown by Hida \cite{hida}, at $\bar{h} - \bar{h}_c \rightarrow 0$,
the two string order parameters vanish as $|\bar{h} - \bar{h}_c|^{1/6}$.

\section{Summary and discussion}

In this paper, we have proposed a nonperturbative approach to describe
the SU(2)$_1$ criticality in the dimerized, weakly-coupled two-leg spin-1/2
ladder. We have shown that this criticality is in fact a quantum critical point
of the effective spin-1 Haldane spin liquid perturbed by the explicit
dimerization. Using the mapping onto a generalized, SO(3)-symmetric,
quantum Ashkin-Teller model and employing a nonlocal duality transformation,
we have derived the low-energy effective
Hamiltonian which represents a lattice $S$=1/2 Heisenberg chain with a small
bond alternation. The SU(2)$_1$ critical point corresponds to the case when
fine tuning of the parameters of the model restores
translational invariance.  With the adopted approach 
we were able to find an asymptotically exact correspondence between the
physical fields of the original spin ladder 
and those characterizing the SU(2)$_1$ criticality at the IR fixed point.

To avoid confusion, let us emphasize that the effective spin-1/2 Heisenberg model
(\ref{eff-final}) should not be misleadingly
associated with the one that corresponds to a {\em strongly} dimerized,
snake looking two-leg ladder. The Hamiltonian $H_{\rm eff}[s_1, s_2]$
was derived in the large-$h$ limit from the generalized, SO(3)-symmetric, 
quantum AT model (\ref{3QIC-rep}) which should be regarded as a 
regularized lattice version of a continuum field theory (\ref{def-WZW-2}),
the latter describing universal properties of a
{\em weakly} dimerized spin-1 chain. No such field-theoretical
description is available if the in-chain staggering amplitude $J'$ 
of the original spin ladder is not small.

The Ising-model description of quantum critical points in two-leg
spin-1/2 ladders can be extended to a more interesting situation where all four
Ising models associated with the triplet and singlet Majorana modes
(see Eqs.~(\ref{ham-maj}) and (\ref{total-ham-cont-fin})) become equally
important. This turns out to be the case for a generalized spin ladder \cite{NT}
\be
H_{\rm gen} = H_{\rm stand}
+ V \sum_n \left( {\bf S}_{1,n} \cdot {\bf S}_{1,n+1}  \right)
\left( {\bf S}_{2,n} \cdot {\bf S}_{2,n+1}  \right) \;, 
\label{gen.lad}
\ee
which, apart from the standard on-rung exchange $J_{\perp}$
present in $H$, Eq.~(\ref{snake}), also includes
a biquadratic interchain interaction. At $J_{\perp} = 0$ the model
(\ref{gen.lad}) is known as the spin-orbital chain\cite{kugel} and has been recently
studied by different groups (see e.g. \cite{pati, cabra, azaria}).
If $J_{\perp} \neq 0$ but $|V|$ is large enough, the model
(\ref{gen.lad})
occurs in a non-Haldane, spontaneously dimerized phase where
the spectrum is entirely incoherent and consists of pairs of topological 
kinks\cite{NT}. When an external longitudinal dimerization of the chains is
included, the generalized ladder is expected to exhibit a pattern of
quantum criticalities presumably richer than that in the standard-ladder
case discussed in Ref.~\cite{snake} and in the present paper. The underlying
SO(3)$\times$Z$_2$ symmetry of the model (\ref{gen.lad}) opens a room
for criticalities with central charge $c = 1/2,1$ and $3/2$, correspoding to
the universality classes of the Ising model and SU(2)$_k$ WZNW models with
$k=1$ and $2$. Recent numerical results\cite{martins} strongly suggest the
appearance of critical points due to the interplay between the
biquadratic interaction $V$ and the explicit dimerization.
This and related problems are presently under investigation.  

%%%%%%%%%%%%%%%%%%%%%%%%%%%%%%%%%%%%%%%%%%%%%%%%%%%%%%%%%%%%%%%%%%%

\bigskip
%%%%%%%%%%%%%%%%%%%%%%%%%%%%%%%%%%%%%%%%%%%%%%%%%%%%%%%%%%%%%%%%%%%
\begin{flushleft}
{\large \bf Acknowledgments}
\end{flushleft}

%{\bf Acknowledgements}
%\medskip

We are grateful to M. Fabrizio, A. O. Gogolin, G. Mussardo,
V. Rittenberg, A. M. Tsvelik and Yu Lu
for fruitful discussions. A.N. is partly supported by the INTAS-Georgia
grant No.97-1340.
%%%%%%%%%%%%%%%%%%%%%%%%%%%%%%%%%%%%%%%%%%%%%%%%%%%%%%%%%%%%%%%%%%%

\appendix
\section{KNK duality transformation}

Consider a model of two noncritical QI chains with an AT-like interchain
interaction
\bea
H_{\rm QI}&=& - \sum_{a=1,2} \sum_{j=1}^N
\left( A_a s^z _{a,j}s^z _{a,j+1} + B_a s^x _{a,j} \right)\nonumber\\
&& -\sum_{j=1}^N \left(A_3 s^z _{1,j}s^z _{1,j+1}s^z _{2,j}s^z _{2,j+1}
+ B_3 s^x _{1,j}s^x _{2,j} \right) \;.
\label{2-QI-chains}
\eea
The KNK transformation \cite{KNK,KT},
which is 
a special combination of spin rotation and duality transformation,
provides a mapping of the model (\ref{2-QI-chains})
onto a quantum spin-1/2 chain. Here we outline
basic steps of this transformation:

(i) duality transformation in the second QI copy:
$$s^x _{2,j}=\nu^z _{2,j-1/2}\nu^z _{2,j+1/2} \;, ~~~~
s^z _{2,j}s^z _{2,j+1}=\nu^x _{2,j+1/2} \;;$$

(ii) reduction of the original ($\{ j \}$) and dual ($\{ j+1/2 \}$) chains
to a single chain via the following identification:
$$
s^{\alpha}_{1,j} \Rightarrow \mu^{\alpha}_{2j} \;, ~~~~
\nu^{\alpha}_{2,j+1/2} \Rightarrow \mu^{\alpha}_{2j+1} \;;
$$

(iii) relabeling the lattice sites, $j\rightarrow j-1/4$, and treating
the sublattice $\{2j + 1/2 \}$ as dual to $\{2j \}$.

(iv) duality transformation:
$$\s^x _n = \mu^z _{n-1/2}\mu^z _{n+1/2} \;, ~~~~
\s^z _n \s^z _{n+1}= \mu^x _{n+1/2} \;;$$

(v) a staggered $\pi$-rotation around the $y$-axis:
$$\s^z _n \rightarrow (-1)^n \s^z _n \;,~~~~
\s^x _n \rightarrow (-1)^n \s^x _n \;;$$

(vi) global $-\pi/2$-rotation around the $x$-axis:
$$
\s^y _n \rightarrow - \s^z _n \;, ~~~~\s^z _n \rightarrow \s^y _n \;.
$$
This brings the Hamiltonian (\ref{2-QI-chains}) to its final form
--- a bond-alternating XYZ $S$=1/2 chain:
\be
H_{\rm XYZ} = 4\sum_{j=1}^N  \sum_{\alpha=x,y,z}\left(
J^{\alpha}_1 T^{\alpha}_{2j} T^{\alpha}_{2j+1}
+ J^{\alpha}_2
T^{\alpha}_{2j-1} T^{\alpha}_{2j}\right) \;, 
\label{xyz-ham}
\ee
where $T^{\alpha}_i = (1/2)\s^{\alpha}_i$ are the spin-1/2 operators,
and
\bea
&&J^x _1 = A_1 \;, ~~~~J^y _1 = A_2 \;, ~~~~ J^z _1 = A_3 \;; \nonumber\\
&&J^x _2 = B_2 \;, ~~~~J^y _2 = B_1 \;, ~~~~ J^z _2 = B_3 \;. \nonumber
\eea

The KNK transformation establishes a
correspondence between the spin operators on the even and odd
lattice sites, $T^{\alpha}_{2j}$ and $T^{\alpha}_{2j+1}$, and products of two
Ising order and disorder operators of the two-chain model
(\ref{2-QI-chains}), $\left(s_1, \nu_1 \right)$ and 
$\left(s_2, \nu_2 \right)$:
\bea
&&2T^x _{2j} = \kappa_1 s^z _{1,j} \nu^z _{2,j+1/2} \;, ~~~~~~~
2T^x _{2j+1} = - \kappa_1 s^z _{1,j+1} \nu^z _{2,j+1/2} \;, \nonumber\\
&&2T^y _{2j} = \kappa_2 \nu^z _{1,j+1/2} s^z _{2,j} \;, ~~~~~~~
2T^y _{2j+1} = - \kappa_2 \nu^z _{1,j+1/2} s^z _{2,j+1} \;, \nonumber\\
&&2T^z _{2j} = \ri \left( \kappa_1 \kappa_2 \right)
\left( s^z _{1,j} s^z _{2,j}\right)  
\left( \nu^z _{1,j+1/2}\nu^z _{2,j+1/2} \right) \;, \nonumber\\
&&2T^z _{2j+1} = -\ri \left( \kappa_1 \kappa_2 \right)
\left( s^z _{1,j+1} s^z _{2,j+1} \right)
\left( \nu^z _{1,j+1/2} \nu^z _{2,j+1/2} \right) \;.
\label{spins-KT}
\eea
Given the standard algebra of the operators $s^z _{1(2),j}$ and $\nu_{1(2),j+1/2}$ 
(see Eqs.~(\ref{Majorana-fusion1}), (\ref{Majorana-fusion2})),
the Klein factors $\kappa_1$, $\kappa_2$ in (\ref{spins-KT})
ensure the correct algebra of the Pauli matrices $2 {\bf T}_n$.

From (\ref{spins-KT}) it follows that
\bea
&&4T^x _{2j}T^x _{2j+1} = -s^z _{1,j}s^z _{1,j+1} \;, ~~~~
4T^y _{2j}T^y _{2j+1} = -s^z _{2,j}s^z _{2,j+1} \;, \nonumber\\
&&4T^z _{2j}T^z _{2j+1} = -s^z _{1,j}s^z _{1,j+1}s^z _{2,j}s^z _{2,j+1} \;;
\label{bilin-KT1}\\
&&4 T^x _{2j-1}T^x _{2j} = - s^x _{2,j} \;, ~~~~
4T^y _{2j-1}T^y _{2j} = -s^x _{1,j} \;, \nonumber\\
&&4 T^z _{2j-1}T^z _{2j} = -s^x _{1,j}s^x _{2,j} \;. 
\label{bilin-KT2}
\eea
Eqs.~(\ref{bilin-KT1}), (\ref{bilin-KT2}) lead to the correspondence
(\ref{su2-inv2}). 
Therefore, the hidden SU(2) symmetry of the following model of two coupled
QI chains  
\bea
H &=& - \sum_{j=1}^N [B (s^x _{1,j} + s^x _{2,j} + s^x _{1,j} s^x _{2,j} )
\nonumber\\
&& + A (s^z _{1,j} s^z _{1,j+1} + s^z _{2,j} s^z _{2,j+1}
+ s^z _{1,j} s^z _{1,j+1} s^z _{2,j} s^z _{2,j+1})]
\label{xxx}
\eea
is encoded in the special Ising structure of the $A$ and $B$ terms 
in the r.h.side of (\ref{xxx}).

\section{Strong-coupling expansion}

In this Appendix, we provide some details concerning the 
unitary transformation of the original
model (\ref{ham-new-rep}) which makes it possible
to derive the effective Hamiltonian $H_{\rm eff}[s_1, s_2]$ in the form
of $1/\bar{h}$ expansion.

It is suitable to rearrange the Hamiltonian in (\ref{ham-new-rep}) in the
following way:
\bea
H_{\rm QI} &=& H + V \;, \nonumber\\
H = H_0 + W \;, && W = W_1 + W_2 \;, 
\eea
where
\bea
&&H_0 = \bar{h} \sum_n \tau^z _n \;, ~~~~ V = - \Delta \sum_n Q_n (s) \tau^x _n \;, 
\nonumber\\
&& W_1 = - \left({\cal J}+K\right) \sum_n \Lambda_{n,n+1}(s)
- K\sum_n R_n (s) \;,\nonumber\\
&&W_2 = \frac{1}{2}\sum_n P_{n,n+1} (s) \left(\tau^z _n - \tau^z _{n+1}\right)^2 \;.  
\label{s-comb}
\eea
In Eqs.~(\ref{s-comb})
\bea
Q_n (s) &=& s^x _{1,n} + s^x _{2,n} + 1 \;, \nonumber\\
R_n (s) &=& s^x _{1,n} + s^x _{2,n} +  s^x _{1,n} s^x _{2,n} \;,\nonumber\\
\Lambda_{n,n+1}(s)  &=& s^z _{1,n}s^z _{1,n+1}+ s^z _{2,n}s^z _{2,n+1}
+s^z _{1,n}s^z _{1,n+1}s^z _{2,n}s^z _{2,n+1} \;, \nonumber\\
P_{n,n+1} (s) &=& {\cal J} s^z _{1,n}s^z _{1,n+1}s^z _{2,n}s^z _{2,n+1}
+K \left(  s^z _{1,n}s^z _{1,n+1} + s^z _{2,n}s^z _{2,n+1} \right) \;.
\eea

Suppose that the eigenvalue problem for $H$ is solved:
$$
H |a\ra = E_a |a\ra \;.
$$
Since $\left[ H_0, W \right] = 0$, any state $|a\ra$ can be represented
as a direct product $|a\ra = |\tau\ra \otimes |s\ra$, where
$|s\ra$ only involve quantum numbers characterizing the
$(s_1,s_2)$ part of the spectrum, while
$$
|\tau\ra = \prod_i ^N |\tau_i \ra
$$
with $|\tau_i\ra = |\up \ra_i , |\down \ra_i$ being eigenstates of 
the operator $\tau^z _i$. 
Under the condition (\ref{s-c.limit}), the spectrum of $H$ coincides in the leading order
with that of $H_0$.  The ground state of $H_0$ 
is a fully polarized state
\be
|0\ra_{\tau} = |\down\down\down...\down\ra_{\tau} \;, 
\label{ground-state}
\ee
while the lowest excited states 
involve one $\tau$-spin flip
\be
|\up\down\down\down...\down\ra_{\tau},~~ 
|\down \up\down\down...\down\ra_{\tau},~~|\down\down\up\down...\down\ra_{\tau} \cdots \;, 
\label{1-flip}
\ee
and have a gap $2\bar{h}$.
With the $s$-degrees of freedom taken into account ($H_0 \rightarrow H_0 + V$), 
the ground state and 1-flip excited
states transform to narrow bands. Our goal is to eliminate $V$ in the lowest order
and project the resulting Hamiltonian onto the state $|0\ra_{\tau}$.

The procedure is standard. Consider a unitary transformation
\bea
\tilde{H}_{\rm QI} &=& e^S H_{\rm QI} e^{-S} \nonumber\\
&=& H + V + \left[S,H \right] + \left[S,V \right]
+ \frac{1}{2} \left[S, \left[S,H\right] \right] + \cdots \;, 
\eea
where $S = -S^{\dagger}$.
Requiring that $V + \left[S,H \right]=0$, one finds that, in the lowest
order in $\Delta/\bar{h}$, the projected Hamiltonian
\be
P_0 \tilde{H}_{\rm QI} P_0 = - 2N\bar{h} + W_1
+\frac{1}{2} P_0\left[S,V \right]P_0
+ O\left[\left(\Delta/\bar{h}\right)^2\right] \;, 
\label{unit-transf-H}
\ee
where
$$
P_0 = |0\ra_{\tau}\la 0 |_{\tau} \;. 
$$
The matrix element of the commutator in
(\ref{unit-transf-H}) is given by
\bea
&&\la a|\left[S,V \right]|b\ra = 
- \sum_c V_{ac}V_{cb}\left( \frac{1}{E_c - E_a} +
 \frac{1}{E_c - E_b}\right) \nonumber\\
&=& - \int_0 ^{\infty} \rd \lambda~ e^{-2h \lambda}
\sum_c \left[\tilde{V}_{ac} (\lambda)V_{cb} + V_{ac}\tilde{V}_{cb}(-\lambda) 
\right] \;, 
\label{int-rep}
\eea
where
$$
\tilde{V}(\lambda) = e^{\lambda W} Ve^{-\lambda W} \;. 
$$
In obtaining (\ref{int-rep}), we assumed
that the states $|a\ra$ and $|b\ra$
are of the form $|0 \ra_{\tau} \otimes |s\ra.$ 
 We also took into account
the fact that the energy differences in
(\ref{int-rep}) are necessarily positive (of the order of $2\bar{h}$)
because the off-diagonal operator $V$ connects $|0\ra_{\tau}$
with 1-flip states (\ref{1-flip}).
Using $1/\bar{h}$ expansion, we obtain:
\be
P_0\left[S,V \right]P_0
=- P_0 \left( \frac{1}{\bar{h}}V^2 + \frac{1}{4\bar{h}^2} \left[ \left[ W,V \right],V \right]
\right)P_0 + O\left(\frac{1}{\bar{h}^3}\right)  \;. 
\label{semi-fin}
\ee
Calculating the commutators in (\ref{semi-fin}) and using formulas
(\ref{bilin-KT1}) and (\ref{bilin-KT2}), one arrives at the results
(\ref{eff-final}), (\ref{J1-J2-J3}) given in the main text.

%%%%%%%%%%%%%%%%%%%%%%%%%%%%%%%%%%%%%%%%%%%%%%%%%%%%%%%%%%%%%%%%%%%%%%%%%%%%%%%%

\end{document}